# Angular Power Spectrum Estimation using High Performance Reconfigurable Computing

Brett Hayes, Robert Brunner
Department of Astronomy, UIUC
{bphayes2|rb}@astro.uiuc.edu

Volodymyr Kindratenko
NCSA, UIUC
kindr@ncsa.uiuc.edu

## Abstract

*Angular power spectra are an important measure of the angular clustering of a given distribution. In Cosmology, they are applied to such vastly different observations as galaxy surveys that cover a fraction of the sky and the Cosmic Microwave Background that covers the entire sky, to obtain fundamental parameters that determine the structure and evolution of the universe. The calculation of an angular power spectrum, however, is complex and the optimization of these calculations is a necessary consideration for current and forthcoming observational surveys. In this work, we present preliminary results of implementing angular power spectrum estimation scheme on a high-performance reconfigurable computing platform.*

## 1. Observational Importance

### 1.1 Cosmic Microwave Background

With the launch of the Wilkinson Microwave Anisotropy Probe (WMAP) in 2001, our understanding of the universe was vastly expanded due to the ability to provide precise measurement of temperature differences in the Cosmic Microwave Background (CMB). The CMB is the radiation that provides one of the earliest views of the universe. The measured angular power spectrum of these extremely small temperature differences is one of the most exceptional results in Cosmology in the last decade [1].

The angular power spectrum of the CMB is a vital tool for understanding the components of our universe at a very early time, just 300,000 years after the Big Bang. By measuring the angular power spectrum of the CMB, a number of Cosmological parameters can be determined, including the density of normal matter, dark matter, and dark energy, as well as the age of the universe.

### 1.2 Galaxy Surveys

Recent galaxy surveys, such as the Sloan Digital Sky Survey (SDSS) and Two-degree Field Galaxy Redshift Survey (2dFGRS), have observed galaxies over enormous cosmic volumes, allowing us to determine the angular power spectrum for structures over large regions of the universe. Power spectra measurements from these large scale surveys are different from CMB spectra, as unlike the CMB, galaxy surveys do not cover the entire sky, and rather than measuring temperature differences, we pixelate the sky and measure galaxy overdensities within the pixels to create a "smooth" distribution.

Galaxies arose later than the CMB, so they probe a different era of the universe. Galaxy clustering is primarily determined by the gravitational attraction between galaxies. However, since we can observe galaxies for most of the age of the universe, it is possible to study the evolution of clustering, and therefore the evolution of the densities of both normal and dark matter throughout time [2].

## 2. Angular Power Spectrum

### 2.1 Calculation

The calculation of an angular power spectrum is not a trivial problem. A full derivation requires maximizing the statistical likelihood of a model fit to the data which involves iteratively adjusting the model to obtain the "best" fit. The solution to this mathematical problem, however, is known, and can be reduced to an analytically simple, yet computationally complex formula.

The first step in computing an angular power spectrum is to pixelize the sky covered by the galaxy survey into $N_p$ pixels, and to quantify the galaxy density within each pixel (or in the case of the CMB, with temperature). From a scientific standpoint, we would like the smallest pixels allowed by the quantity of data, and to measure as many components of the spectrum, which is done in the bandpowers $C_l$, as possible. From these $N_p$ pixels, we construct a covariance matrix ($\Delta\Delta^T$) of size $N_p^2$. By using bandpowers, we construct a similar matrix, called the signal matrix denoted by C, and perform a series of matrix multiplications and inversions to make the signal matrix, as defined by the bandpowers, iteratively converge to the covariance matrix given by our data. The following two equations describe the iterative procedure. The $C_l$ are used to construct C, and these equations produce the next iteration values of $C_l$:

$$F_{ll'} = \frac{1}{2} Tr(C^{-1} \frac{\partial C}{\partial C_l} C^{-1} \frac{\partial C}{\partial C_{l'}})$$

$$\delta C_l = \frac{1}{2} F_{ll'}^{-1} Tr[(\Delta\Delta^T - C)(C^{-1} \frac{\partial C}{\partial C_{l'}} C^{-1})]$$

## 2.2 Technical Difficulty

Though approximations can be made to improve the performance of the calculation, in this poster we present the results of the full, "brute force" calculation. In this approach, we want $N_p$ and $N_B$ to be as large as possible. Unfortunately, each iteration of the algorithm requires $2N_B^2 + 3N_B$ matrix multiplications, each of which scale as $N_p^3$. For the data sets that we are analyzing we use a basic resolution of $N_p \sim 5{,}000$ and $N_B \sim 10$, and ideally both of these numbers would be higher given sufficient computational capacity.

## 2.3 The Role of Reconfigurable Computing

Naturally, the first place to start reducing the computational time involved is to attempt to optimize the operation that scales as $N_p^3$: the matrix multiplication. Our current area of research is to attempt to do the matrix multiplication on Field Programmable Gate Array (FPGA)-based platform which can be configured to our specific application. Our platform of choice is SRC-6 MAP Series E processor.

FPGAs excel in speeding up simple repetitive calculations; however, our preliminary results indicate that they may not be well suited to our particular problem. While the large matrix multiplication can be implemented its performance is limited by the off-chip memory bandwidth. Since SRC-6 MAP Series E processor provides eight on-board memory banks that allow only six read/writes to memory per clock cycle, only a fraction of the MAP processor's peak floating point performance can be effectively utilized.

## 2.4 Parallelization

Another method of speeding the calculation is to parallelize the $\sim N_B^2$ matrix multiplications that can be done simultaneously. This has the advantage of being completely independent of the resolution we choose, it only depends on the number of bandpowers that we calculate. With more bandpowers, we would require more processors, while maintaining constant calculation time.

This process is suitable for traditional supercomputers. Each matrix multiplication, however, still requires the same time for each calculation, and this process cannot be easily parallelized. Though we can reduce every iteration to several sequential matrix multiplications, we would still like to speed up the matrix multiplication itself.

## 3. Conclusion

For today's data sets, the calculation of the angular power spectrum can push the limits of our computational ability. Using reconfigurable computing technologies can perhaps speed the calculation by a factor of a few. Parallelization can decrease the computation time by factors of dozens to hundreds, and hopefully make these calculations faster so that we can obtain more scientific results from the available data.

In addition to angular power spectra, many other similar calculations exist, such as the three-dimensional power spectrum, spherical harmonic expansions, higher order power spectra (bispectrum, trispectrum, etc.), and correlation functions that have already seen great improvement using reconfigurable computing. We anticipate that all of these measurements will benefit from easier and faster calculation, and we are working to provide direct comparisons of the different reconfigurable computing technologies to these Cosmological measurements.